\documentclass[prper,aps,reprint,twocolumn,groupedaddress,floats,showpacs,final,superscriptaddres]{revtex4-1}
\usepackage[latin3]{inputenc}
\usepackage[makeroom]{cancel}
\usepackage{graphicx}
\usepackage{amsmath}
\usepackage{amsfonts}
\usepackage{amssymb}
\usepackage{color}
\usepackage[left=2cm,right=2cm,top=2cm,bottom=2cm]{geometry}

\usepackage{dcolumn}
\usepackage{bm}
\usepackage{simplewick}
\usepackage{array}
\usepackage{appendix}

\RequirePackage[
   hyperindex,colorlinks,bookmarksnumbered,
   plainpages=true,pdfstartview=FitH]{hyperref}
\hypersetup{linkcolor=blue,urlcolor=blue,citecolor=blue}
\usepackage{hyperref}

\definecolor{purple}{rgb}{0.5,0,0.6}

\usepackage{ulem}
\renewcommand{\emph}[1]{\textit{#1}}
\definecolor{darkblue}{rgb}{0,0,0.5}
\definecolor{darkgreen}{rgb}{0,0.5,0}
\definecolor{darkred}{rgb}{.7,0,0}
\definecolor{purple}{rgb}{0.5,0,0.6}
\definecolor{orange}{rgb}{1,0.5,0}
\definecolor{grey}{rgb}{.6,.6,.6}
\definecolor{lightpink}{rgb}{1,0.7,0.75}
\definecolor{pink}{rgb}{1,0.4,0.58}
\definecolor{deeppink}{rgb}{1,0.08,0.58}

\renewcommand{\emph}[1]{\textit{#1}}

\begin{document}

\date{\today}
\title{Tunable RKKY interaction in a double quantum dot nanoelectromechanical device}

\author{A. V. Parafilo}
\author{M. N. Kiselev}
\affiliation{The Abdus Salam International Centre for Theoretical
Physics, Strada Costiera 11, I-34151 Trieste, Italy}


\date{\today}

\begin{abstract}
We propose a realization of mechanically tunable Ruderman-Kittel-Kasuya-Yosida interaction in a double quantum dot nanoelectromechanical device. The coupling
between spins of two quantum dots suspended above a metallic plate
is mediated by conduction electrons. We show that the spin-mechanical 
interaction can be driven by a slow modulation of charge density in the metallic plate.
{\color{black} We propose to use  St\"uckelberg oscillations as a sensitive tool for detection of the
spin and charge states of the coupled quantum dots.} {Theory of mechanical back action induced by a dynamical spin-spin interaction is discussed. }
\end{abstract}
\maketitle
\section{Introduction}
Recent progress in theory and experiment on nanometer-sized devices 
sheds light on
a significant role of the spin degree of freedom (see Refs \cite{Rugar2004}-\cite{material})
in opto- and electromechanical systems.
{\color{black} The nanomechanical {\it quantum - classical hybrid} systems \cite{cleland}-\cite{optomech} are important for both fundamental research and applications  \cite{vdzant}.
{\color{black} The range of problems addressed by the nanomechanics varies from designing new tools for a quantum information processing to a development of highly sensitive methods
of, for example, mass, force and current  detection in metrology  \cite{cleland},\cite{vdzant}. 
While the nano-optomechanics  is dealing with coupling of light with mechanical degrees of freedom \cite{optomech}, the nanoelectromechanics (NEM) works with mechanically nanomachined electrons \cite{vdzant}. Several new directions of NEM, in particular those which are focused
on an investigation of mechanical systems coupled to the spins (spintromechanics \cite{rev} and optomagnonics \cite{optomagnons}) emerged recently thanks to a progress in both experiment \cite{NVcenters}, theory \cite{optomagnons} and material science \cite{material}.
The ability to manipulate nanoelectromechanical systems via electron's spins leads to a variety of new phenomena \cite{flens}-\cite{low}. Since typical mechanical displacements of NEM devices are in a range from angstroms to nanometers, its detection requires utilization of very sensitive methods. Quantum interferometry \cite{schevch} provides
one of such sensitive tools \cite{eva},\cite{ch}. In most cases measurements of a back action induced by the quantum electron spin \cite{nori1} and charge system \cite{okazaki16},\cite{nori4} onto a mechanical resonator gives yet another sensitive tool for measuring the out-of-equilibrium properties of the quantum system operating in many cases in a regime of strong electron-electron interaction and/or resonance scattering, see examples in Ref.\cite{rev}.}

One of the most intriguing examples of spin-related physics
in NEMS is the Kondo effect in shuttling devices \cite{kiselev1},\cite{kiselev2}. Kondo physics in quantum dots (QD) \cite{glazman} manifested itself as a many-body effect
associated with the creation of a cooperative singlet state composed of conduction electrons in the leads and a localized QD spin $S$$=$$1$$/$$2$.
Complete screening {\color{black} of a spin impurity in the QD} occurs at temperatures well below the Kondo temperature $T_{\rm K}$, the typical energy scale of the interaction. Formation of a Kondo singlet is accompanied by the saturation of the 
nano-device's electric conductance at the  unitary limit {\color{black} $2e^2/h$}. Mechanical motion of the QD results in the appearance of a time dependency of $T_{\rm K}$, which allows us to employ this effect as a dynamical probe of the  Kondo cloud \cite{kiselev2}. Quantum engineering of  NEM-QD devices opens
a possibility for investigating competing interactions and emergent symmetries
in the presence of resonance electron scattering, for example:
two channel Kondo effect in a side-coupled QD \cite{potok}, two impurity Kondo effect in parallel- and serially-coupled double quantum dot (DQD) \cite{craig},\cite{rew}, or SU(4) Kondo effect in a single-wall carbon nanotube based QD \cite{su4}, \cite{su42}. 

To demonstrate back action based on spin exchange in a mechanical resonator we concentrate on studying a parallel DQD system with spin-spin coupling controlled by its nanomechanical motion. In the regime of resonant scattering of mobile electrons on localized spins, the two-impurity Kondo model arises \cite{krishna}.
It is well known (see, e.g.,
Refs. \cite{krishna},\cite{jones})  that at temperatures 
$T$$<$$T_{\rm K}$, mobile electrons "screen" each impurity independently. 
It happens, however only if the spin-spin interaction mediated by conduction electrons,
aka Ruderman-Kittel-Kasuya-Yosida (RKKY) interaction \cite{RKKY} is sufficiently small.
The system behaves as two independent Kondo impurities under condition  $|I_{\rm RKKY}|$$\lesssim $$2.2 T_{\rm K} $ \cite{crit}, otherwise the two impurities' spins are effectively locked into a singlet or triplet state depending on the sign of the RKKY interaction $I_{\rm RKKY}$. In recent experiments  the RKKY interaction carried by the electrons in metallic grain \cite{craig} or a nanowire \cite{sasaki} was used as additional "tools" for manipulating the Kondo effect in DQD systems, see, e.g., Ref. \cite{twodot}. However, as shown in theory \cite{glva}, \cite{rosa}, the controllability of these systems, e.g., tunability of RKKY interaction is very much limited by the device fabrication processes.

In this {\color{black}paper} we show theoretically that in a suspended DQD NEM hybrid device placed near a metallic plate (back gate), the coupling between mechanics and 
electron spin-1/2
localized in the QDs can be mediated by the RKKY interaction. We investigate two NEM subsystems suspended above a two-dimensional electron gas (2DEG) \cite{foot}, see Fig.~\ref{Fig1}(a). Each subsystem consists of a source  and a drain  bridged by a vibrating QD.
A molecule attached to the leads by the van-der-Waals interaction or another vibrating nanowire with strong size quantization, e.g. carbon nanotube with length less than $1\,{\color{black}\rm \mu m}$
serves as a prototype QD device. We assume that the number of electrons on each QD is odd and consider QDs as mobile spin quantum impurities. 

The possibility of electrons to tunnel from QDs to the 2DEG and back leads to  effective exchange (RKKY) interaction between the QDs with spins $S_1$ and $S_2$,
$H_{\rm eff}$$= $$I_{\rm RKKY}(R){\bf S}_1{\bf S}_2$.  The RKKY interaction {\color{black} mediated by conducting electrons} is an oscillating function of the distance $R$ between the localized spins, 
$I_{\rm RKKY}$$\propto $$\cos(2k_{ F} R$$-$$\pi/2(D$$+$$1))/R^D$ (see Ref.\cite{aristov}), 
where $k_F$ is the electron momentum at the Fermi surface and $D$ is the dimension of the metallic reservoir. This interaction provides an implicit coupling
between mechanical and spin degrees of freedom originating from
the time-dependent deflection of the vibrating QDs,  see Fig.~\ref{Fig1}(b). 

The sign of exchange interaction determines the ground state of the  DQDs spin configuration. Namely, for the distance $R(t)$ such that $I_{\rm RKKY}(R(t))$$>$$0$, the interaction between spins is antiferromagnetic and the corresponding ground state of the system is a singlet (the total spin $S$$=$$0$). The ferromagnetic RKKY interaction $I_{\rm RKKY}(R(t))$$<$$0$
facilitates formation of the triplet $S$$=$$1$ ground state. {\color{black}As a result, the potential energy associated with the RKKY exchange interaction gives rise to an additional
displacement-dependent force. This force acts on the mechanical resonator being sensitive to the spin configuration of the DQD NEM device}.

{\color{black} The paper is organized as follows: Section II is devoted to formulation of the model describing a driven double-quantum dot nanoelectromechanical device. We present a short derivation of the mechanically nanomachined effective two-impurity Kondo model. The RKKY induced back action in mechanical subsystem and the dynamics of the QD oscillations are discussed in Sec. III. The analysis of the St\"uckelberg interference pattern is presented in Sec. IV. The summary and discussions are given in Sec. V. }

\begin{figure}
\centering
\includegraphics[width=1.\columnwidth]{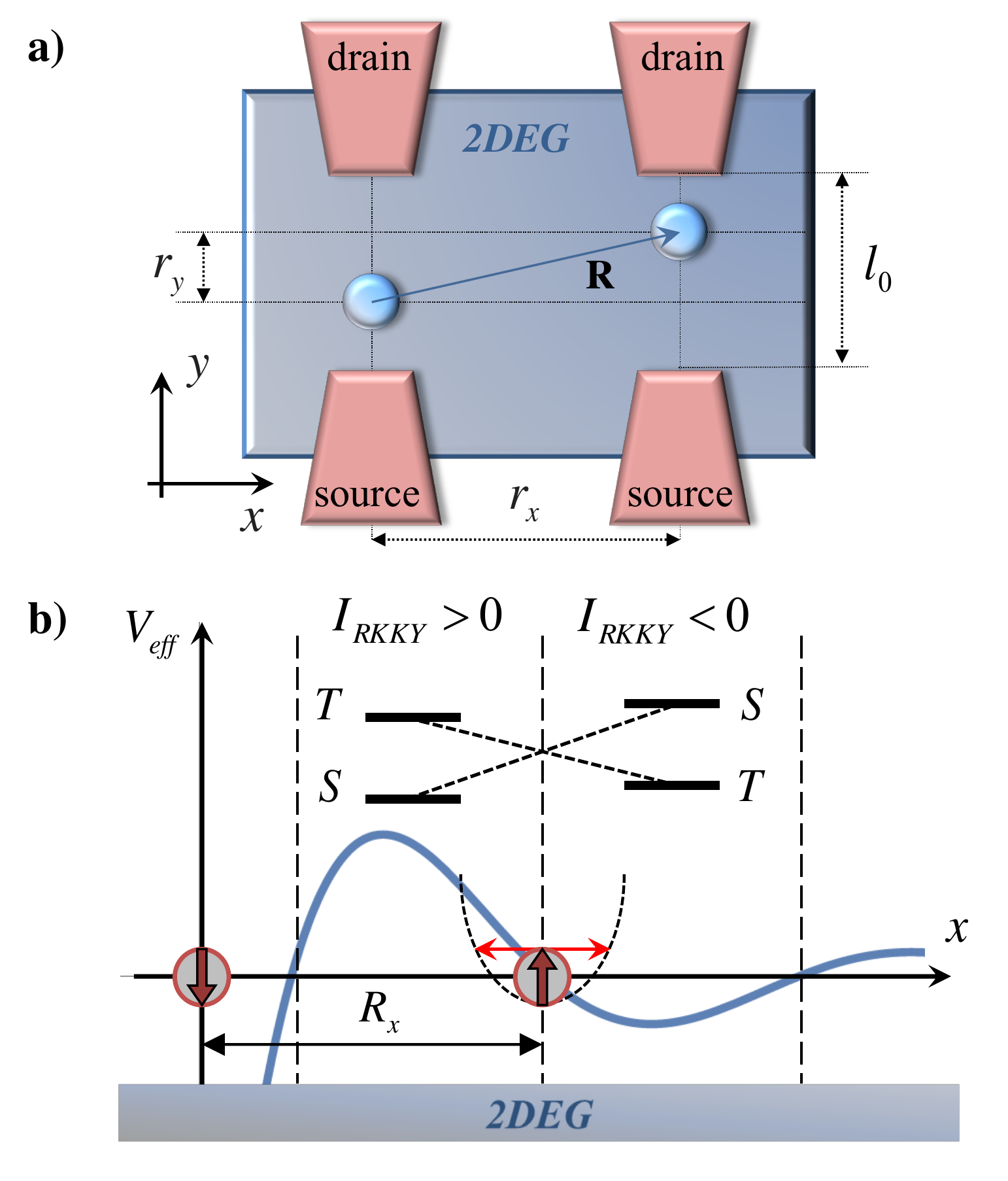}
\caption{(Color online) (a) Mobile double quantum dot NEM device: Each quantum dot (light blue circle) is sandwiched between its own source-drain electrodes (pink). The quantum dots are suspended above a metallic plate (gray) formed by 2DEG. The dots are attached to the
mechanical oscillators (not shown) and move {\color{black} in the $x$ direction} parallel to the metallic plate (motion is excited by external gates not shown on the picture).
The time-dependent distance between QD is denoted by ${\bf  R}(t)$; $r_y$ and $r_x$ are Cartesian projections of the DQDs equilibrium (rest) position, $l_0$ is a length of mechanical oscillator.
(b) Sketch of RKKY interaction between two local spins attached to mechanical resonator $V_{\rm eff}$$=$$I_{\rm RKKY}(R)\langle{\bf S}_1{\bf S}_2\rangle$, where sign of interaction is determined by $R$$=$$|{\bf R}|$.
The ground state of the two-localized-electron system is singlet (S) if $I_{\rm RKKY}(R)$$>$$0$
and triplet (T) otherwise. If the mechanical system is fine tuned to the nodal points of the curve, the RKKY interaction changes its sign along the trajectory. The potential energy
associated with RKKY interaction gives rise to an additional force acting onto the mechanical system.} \label{Fig1}
\end{figure}

\section{Model}
The Hamiltonian of the mobile DQD system reads:
\begin{equation}\label{Hamil}
H=H_{\rm DQD}+H_{\rm leads}+H_{\rm tun}+H_{\rm vib},
\end{equation}
where the first term characterizes the DQD
\begin{eqnarray}\label{dot}
H_{\rm DQD}=
\sum_{i=1,2,\sigma,\sigma'}\left[\varepsilon_i\hat n_{i,\sigma}+U\hat n_{i,\uparrow}\hat n_{i,\downarrow}\right]+U_{12}\hat n_{1,\sigma}\hat n_{2,\sigma'}.\;\;\;\;\;\;
\end{eqnarray}
Here $\varepsilon_i$ is an electron energy level on the $i$th QD, $\hat n_{i,\sigma}$$=$$\hat d_{i,\sigma}^{\dag}\hat d_{i,\sigma}$ is a density operator of electron with spin projection $\sigma=\uparrow,\downarrow$ and $\hat d_{i,\sigma}(\hat d_{i,\sigma}^{\dag})$ is its annihilation (creation) operator, $U$ is the Coulomb interaction between electrons in each QD, and $U_{12}$ is 
a capacitive coupling
between QDs. We assume that each QD is singly occupied and the system is in the particle-hole symmetric regime, $\varepsilon_i$$=$$-U/2$. $H_{\rm leads}$ in Eq. (\ref{Hamil}) describes the electrons in the leads and metallic plate, 
\begin{eqnarray}\label{lead}
H_{\rm leads}=\sum_{\alpha}\sum_{{ k}, i,\sigma}\xi^i_{ k\alpha}\hat c_{{ k}\sigma\alpha}^{i\dag}\hat c_{{ k}\sigma\alpha}^i+\sum_{{\bf k}\sigma}\epsilon(t)\hat a^{\dag}_{{\bf k}\sigma}\hat a_{{\bf k}\sigma},
\end{eqnarray}
where operators $\hat c_{{ k}\sigma\alpha}^i(\hat c_{{ k}\sigma\alpha}^{i\dag})$ and  $\hat a_{{\bf k}\sigma}(\hat a_{{\bf k}\sigma}^{\dag})$ denote electron annihilation (creation)   in $(i, \alpha)$th  electrode and 2DEG (here $\alpha$ stands for source/drain). Their excitation energies $\xi^i_{{ k}\alpha}$$=$$\varepsilon^i_{{ k}\alpha}$$-$$\mu^i_{\alpha}$ and $\epsilon(t)$$=$$\epsilon_{\bf k}$$-$$\mu(t)$ 
are counted from
the chemical potentials $\mu^i_{\alpha}$, $\mu(t)$. Note that we consider the case of a time-dependent chemical potential and density of the electrons in the 2DEG, $\mu(t)$$=$$\epsilon_F$$+$$eV\sin(\Omega t)$ (periodic modulations with a frequency $\Omega$), while the amplitude of its modulation {\color{black} is limited by a condition} $|eV|$$<$$U/2$ to avoid electron exchange between the QDs and 2DEG. {\color{black} In addition we assume an adiabatic drive (see discussion below)}.

The electron tunneling between leads and QD is described by a standard tunnel Hamiltonian
\begin{eqnarray}\label{tunn}
&&H_{\rm tun}=\sum_{k,\alpha,\sigma,i}(t_{ki,\alpha}\hat c^{i\dag}_{k\alpha\sigma}\hat d_{i,\sigma}+h.c.)\nonumber \\
&&+\sum_{{\bf k},i,\sigma}(\gamma_{{\bf k}i} e^{i{\bf k}{\bf R}_i(t)}\hat a^{\dag}_{{\bf k}\sigma}\hat d_{i,\sigma}+h.c.),
\end{eqnarray}
where $t_{ki,\alpha}, \gamma_{{\bf k}i}$ are corresponding tunnel matrix elements and ${\bf R}_i$ denotes a momentary position of the $i$th QD. 

Oscillations of the $i$th "impurity" with frequency $\omega_i$ are described by a harmonic oscillator Hamiltonian  $H_{\rm vib}$$=$$\sum_i(\hat p^2_i/2M$$+$$M \omega_i^2 \hat x_i^2/2)$, where $\hat p_i$ and $\hat x_i$ are momentum and displacement of the $i$th QD. We assume equal masses $M$ of the oscillators. We point out that QD vibrations take place along the $x$ direction in the plane parallel to the 2DEG, see Fig.~\ref{Fig1}. In the following, we assume that QD' displacements are large compared to the zero-point motion amplitude $x_0$, $\hat x_i$$\gg $$x_0$$=$$\sqrt{\hbar/M\omega_i}$, which allows us to consider $\hat x_i, \hat p_i$ as classical variables.

We map the above model Eqs.(\ref{Hamil})-(\ref{tunn}) onto a two-impurity Kondo problem \cite{jones} by using a time-dependent Schrieffer-Wolff transformation \cite{kaminski}. This procedure is legitimate as long as the number of electrons occupying each QD is odd, which requires fulfillment of the condition {\color{black} $|U|/2$$\gg $$\Gamma^i_{\gamma}$$+$$\sum_{\alpha} \Gamma^i_{\alpha}$}. Here  
$\Gamma^i_{\alpha}$$=$$2\pi \nu |t_{ki,\alpha}|^2$ and 
$\Gamma^i_{\gamma}$$=$$2\pi N_{\rm 2D} |\gamma_{{\bf k}i}|^2$ are the tunneling rates in the $(i,\alpha)$th electrode and 2DEG correspondingly, while $\nu$ and $N_{\rm 2D}$ are the densities of states at the Fermi level. 
The effective Kondo Hamiltonian in adiabatic approximation, {\color{black} $\{\hbar\Omega, \hbar k_F |\dot {\bf R}_i|\}$$\ll$$|U|/2$} \cite{kaminski}, reads
\begin{eqnarray}\label{model}
&&H_{\rm K}=\sum_{{\bf k}\sigma}{\color{black}\epsilon(t)}\hat a^{\dag}_{{\bf k}\sigma}\hat a_{{\bf k}\sigma}+\sum_{{\bf k},{\bf k}';i}{\rm J}(t)e^{i({\bf k}'-{\bf k}){\bf R}_i(t)}{\bf s}\cdot {\bf S}_i+\;\;\\
&& \sum_{{ k}\sigma\alpha;i}\xi^i_{{ k}\alpha}\hat c_{{ k}\sigma\alpha}^{i\dag}\hat c_{{ k}\sigma\alpha}^{i}+\frac{1}{2}\sum_{{ k}{ k}', \alpha\alpha'}\sum_{\sigma\sigma'; i}{\rm J}_{\alpha\alpha'}^i{\bf S}_i\cdot \hat  c^{i\dag}_{{ k}'\sigma'\alpha'}{\color{black}{\bm \sigma}^{\sigma'\sigma}}\hat c^i_{{ k}\sigma\alpha}.\nonumber
\end{eqnarray}
Here {\color{black}${\bf s}$$=$$\hat a_{{\bf k}'\sigma'}^{\dag}{\bm\sigma}^{\sigma'\sigma} \hat a_{{\bf k}\sigma}/2$ and ${\bf S}_i$} are the spin operators of the electrons in the 2DEG and {\color{black} of} $i$th QD, {\color{black}${\bm\sigma}^{\sigma'\sigma}$} are the Pauli matrices, ${\rm J}(t)$$=$${\rm J}_0$$+$${\rm J}_1[(eV\sin(\Omega t)$$-$$({\bf k}$$+$${\bf k}')\dot{{\bf R}}_i/2)]$ is the exchange interaction between the $i$th "impurity" and the 2DEG,{\color{black}
\begin{eqnarray}\label{exch}
{\rm J}_0=\frac{2|\gamma_{{\bf k}i}|^2U}{E_{12}(U-E_{12})} \quad ,\quad {\rm J}_1=\frac{{\rm J}_0^2}{2|\gamma_{{\bf k}i}|^2}\frac{U-2E_{12}}{U}.
\end{eqnarray}
Here $E_{12}$$\equiv$$U$$/$$2$$-$$U_{12}$ and
${\rm J}_{\alpha\alpha'}^i$$=$$2t_{k'\alpha'}^it^{i\ast}_{k\alpha}U$$/$$E_{12}$$($$U$$-$$E_{12}$$)$}
 is the exchange constant between the $i$th localized  spin and electrons in the leads.  In Eq. (\ref{model}) we omit irrelevant terms of two sorts: (i) responsible for electron scattering without spin flip and (ii) accounting for spin-flip processes during electron scattering from {\color{black} source/drain} electrodes to the 2DEG and vice versa. Below we restrict ourself to the question of how RKKY interaction in the DQD-NEM system affects the vibrational degree of freedom. 
For simplicity we assume $|{\rm J}_0|$$\gg $$T_{\rm K}^i$ and ignore Kondo physics {\color{black} (two different Kondo temperatures \cite{twodot},\cite{glva} generically arise in  the four-terminal setup, Fig.~\ref{Fig1}: Two pairs of contacts provide two independent Fermi seas for the quantum impurities).}
We neglect also all effects associated with current flow through the system by considering zero source-drain bias. {\color{black}The out-of-equilibrium effects associated with finite current
and effects of thermal noise due to equilibrium current fluctuations will be considered elsewhere \cite{tbp1}.}

\section{Back-action}
We analyze the dynamics of the $i$th QD characterized by the amplitude of its fundamental 
vibrational mode $x_i(t)$, whose time evolution is described by Newton's equation 
\begin{equation}\label{newton}
 \ddot x_i+\gamma \dot x_i + \omega_i^2 x_i=-\frac{1}{M}\frac{\partial}{\partial x_i}V_{\rm \!ef\!f}(|{\bf R}_1-{\bf R}_2|, t),
 \end{equation}
here $\gamma$ is a phenomenological damping and $V_{\rm \!ef\!f}$ is an effective exchange interaction, which in adiabatic approximation ($\hbar \Omega \ll\epsilon_F $) can be obtained from the linear response theory \cite{vignale}. 
{\color{black} To derive an effective potential for the RKKY interaction between spins located at different wires we integrate out completely all states of 2DEG at the conduction plate (see Fig.~\ref{Fig1}). As a result, the effective RKKY potential is given by the real part of the density-density correlation function \cite{kittel} of the 2DEG 
\cite{footnote1}. The imaginary part of this correlation function contributes to the damping of the mechanical subsystem.}
Performing a Fourier transform of $V_{\rm \!ef\!f}$ and treating the second term in Eq.(\ref{model}) as a perturbation in {\color{black} $|\gamma_{{\bf k}i}|^2/(U\epsilon_F)\ll 1$}  we obtain
\begin{eqnarray}\label{interaction}
&& V_{\rm \!ef\!f}=\int d\omega e^{-i \omega t}\widetilde{V}_{\rm \!ef\!f}({\bf R}, \omega)\left(1+\frac{{\rm J_1}{\color{black}eV}}{2i{\rm J_0}}\sum_{\kappa=\pm}\kappa\delta(\omega+\kappa\Omega)\right),\nonumber
\\&&\widetilde{V}_{\rm \!ef\!f}(|{\bf R}|, \omega)=\frac{{\rm J}_0^2}{2}\sum_{{\bf k},{\bf q};i\neq j}\langle{\bf S}_i{\bf S}_j\rangle \frac{e^{-i{\bf q}{\bf R}}(f_{{\bf k}}-f_{{\bf k}+{\bf q}})}{\tilde\omega_i+{\color{black}\epsilon}_{{\bf k}}-{\color{black}\epsilon}_{{\bf k}+{\bf q}}+i0^{+}},\label{responce}
\end{eqnarray}
where {${\bf R}$$=$${\bf R}_1$$-$${\bf R}_2$}, {\color{black}$\delta(\omega)$ is the Dirac delta-function}, $\tilde\omega_i$$=$$\omega$$-$${\bf q}\dot{{\bf R}}_i$,  and  $f_{{\bf k}}$ is the Fermi distribution function. 
The real part of the effective potential is the RKKY interaction between two "impurities", while its imaginary part is the spectral function of electron-hole excitations in 2DEG.
The imaginary part of $\widetilde V_{\rm \!ef\!f}$ is in general related to
the damping mechanisms associated with, e.g., the  creation of collective plasmon or/and magnon excitations or tunneling into the leads. We neglect such contributions by assuming the adiabaticity condition.  
The real part of $\widetilde{V}_{\rm eff}(|{\bf R}|,0)$ at zero temperature (see, e.g., Refs. \cite{aristov},\cite{klein}) 
is given by:
\begin{eqnarray}\label{integr}
&&V_{\rm eff}(R)=\frac{{\rm J}_0^2}{2}\langle{\bf S}_1{\bf S}_2\rangle A^2 N_{\rm 2D}k_F^{\ast 2}\cdot \nonumber \\&&\left[J_0(k_F^{\ast}R)N_0(k_F^{\ast}R)+J_1(k_F^{\ast}R)N_1(k_F^{\ast}R)\right],
\end{eqnarray}
where $A$ is the confinement area of the 2DEG {\color{black} (back gate)}, $k_F^{\ast}$$\approx $$k_F$$+$$(eV/2\hbar v_F)\sin(\Omega t)$ is a time-dependent Fermi wave vector,
and $J_{m}(z), N_{m}(z)$ are the Bessel functions of the first and the second kind. For simplicity, we consider only the long-distance limit of Eq.~(\ref{integr}) at large $k_F R$, which corresponds to a power-law 
$-\sin(2k_F^{\ast}R)/R^2$ RKKY asymptote \cite{com1}. Here the distance between "impurities" is $R$$=$$\sqrt{(r_x+x_1-x_2)^2+r_y^2}$  and 
$r_x$, $r_y$ are $x$-$y$ projections of ${\bf R}$
at the equilibrium position, see Fig.~\ref{Fig1}(a).
Let us assume that the averaged distance between the vibrating QDs 
$R_0$$=$$\sqrt{r_x^2+r_y^2}\;$ is close to the distance at which the RKKY interaction changes sign, $2k_F R_0$$\approx $$\pi n$ (where $n$ is an integer). Then, by expanding the oscillating RKKY function with respect to the small parameters ($eV/\epsilon_F, r_x(x_1$$-$$x_2)/R_0^2$$\ll $$1$) and substituting it into Eq.~(\ref{newton}) one can obtain the system of coupled equations of motion for the displacement of the $i$th QD
\begin{eqnarray}\label{EOM1}
&&\ddot x_i+\gamma\dot x_i+\omega^2_i x_i=\pm\alpha_0 \left(1+\frac{2r_y^2}{R_0^2}\frac{x_1-x_2}{r_x}\right)\times\nonumber \\
&&\left(1+eV\left(\frac{1}{2\epsilon_F}+\frac{{\rm J}_1}{\rm{J}_0}\right)\sin(\Omega t)\right),
\end{eqnarray}
\noindent
where $"\!+\!"$ is for $i$$=$$1$, $"\!-\!"$ for $i$$=$$2$, {\color{black} and $\alpha_0$$=$${\rm J}_0^2\langle{\bf S}_1{\bf S}_2\rangle N_{\rm 2D}(A^2k_Fr_x/2\pi M R_0^3)$}. The RKKY interaction results in three different forces in the r.h.s. of Eq.~(\ref{EOM1}) which (i) lead to the renormalization of the $i$th QD equilibrium position, (ii) create a time-dependent force proportional to $\sin(\Omega t)$, and 
(iii) result in energy transferring between the two oscillating QDs (beating). In particular, the RKKY interaction between spins provides the coupling between the QDs and the mechanical subsystems. Such spin-mechanical coupling 
can be easily extended to the quantum limit.
Expansion of $\sin(2k_FR)$ around the equilibrium interdot distance $\pi n/2k_F$ and quantization of the QD displacement field leads to the spin-mechanical interaction Hamiltonian: $H_{\rm int}$$\sim$$ \lambda {\bf S}_1{\bf S}_2 ({\color{black}\hat b+\hat b^{\dag}})/\sqrt{2}$, where 
$\lambda$$=$$(-1)^n{\rm J_0^2}N_{\rm 2D}k_F{A}^2r_xx_0/R_0^3$ is the spin-phonon coupling, 
{\color{black}$\hat b,\hat b^{\dag}$} are boson operators of vibrational quanta.

We rewrite the EOM (\ref{EOM1}) in terms of a normal (i) in-phase, $x_1$$+$$x_2$, and (ii) out-of-phase, $x_1$$-$$x_2$, modes. 
The first term in the r.h.s of Eq. (\ref{EOM1}) is eliminated by redefinition of the  "impurities" initial deflections $x_i $$\mp$$ \alpha_0/\omega_i^2$$\rightarrow$$ x_i${\color{black}.}
We assume that the QD eigenfrequencies {\color{black}$\omega_{1,2}$$=$$\omega_0\pm\delta\omega$ differ by a small value $\delta\omega$$\ll$$\omega_0$}. In addition, we introduce a dimensionless time $\tau$$=$$\omega_0 t$ and dimensionless normal mode displacements in units of the length of the nanowires $l_0$: $\varphi$$=$$(x_1+x_2)/l_0$, $\phi$$=$$(x_1-x_2)/l_0$. We denote
\begin{eqnarray}\label{def}
\Delta=\frac{2\delta\omega}{\omega_0}\quad, \quad \omega_d=\frac{\Omega}{\omega_0}\quad,\quad \tilde{\alpha}_0=\frac{2\alpha_0}{\omega_0^2},
\end{eqnarray}
and introduce dimensionless force and frequency shifts
$$
{\color{black}F}=\frac{\tilde{\alpha}_0eV}{l_0}\frac{{\rm J_1}}{{\rm J_0}},\;\; \alpha_1=2\frac{\tilde{\alpha}_0}{r_x}\left(\frac{r_y}{R_0}\right)^2,\;\;\alpha_2=\frac{l_0 {\color{black}F}}{r_x}\left(\frac{r_y}{R_0}\right)^2,
$$
The {\color{black} coupled mechanical equations of motion} (\ref{EOM1}) in dimensionless notations are given by
\begin{eqnarray}\label{nm}
&&\ddot \varphi + \dot \varphi/Q+ \varphi= -\Delta\cdot \phi,\\
&&\ddot \phi +\dot\phi/Q+[1-\alpha_1-\alpha_2\sin(\omega_d \tau)]\phi =-\Delta\cdot \varphi +{\color{black}F}\sin(\omega_d \tau),\nonumber
\end{eqnarray}
{\color{black} where $Q$ is a quality factor.} As a result, the {\color{black} equation of motion} for the out-of-phase mode $\phi$ describes a parametric oscillator subject to an external time-dependent driving force coupled to a 
nondriven oscillator, associated with the in-phase mode $\varphi$. Notice that the frequency shift for the in-phase mode is negligible compared to the shift for the out-of-phase mode which consists of a time-independent part $\alpha_1$ and a part $\propto\alpha_2$ periodic in time.

The coupling constant $\alpha_2$ in (\ref{nm}) can be estimated considering realistic parameters for a typical 2DEG: $\epsilon_F$$\sim $$10\, {\color{black}\rm meV}$, 
 $k_F$$\sim $$10^6\, {\color{black}\rm cm}^{-1}$. 
 Besides, without loss of generality we assume ${\rm J}_0$$\sim $$10\,{\color{black}\rm K}$ and   
 $eV/\epsilon_F $$\sim $$ 0.1$
and consider a carbon nanotube as an example of vibrating nanowire 
($\omega_0 $$\sim $$100\, {\color{black}\rm MHz}$ is a {\color{black} fundamental} frequency of {\color{black} carbon nanotube's} bending modes, $x_0 $$\sim $$10^{-9}\,{\color{black}\rm cm}$ is the amplitude of zero-point oscillations). Furthermore, taking $R_0 $$\sim $$r_y $$\sim $$10^{-6}\,{\color{black}\rm cm}$ and considering the QDs charging energies in the range from $1\, {\color{black}\rm K}$ to $10\, {\color{black}\rm  K}$ we obtain:
$$
\alpha_2\sim \frac{{\rm J}_0}{\epsilon_F}\frac{{\rm J}_0}{\hbar \omega_0}\frac{1}{k_FR_0}\left(\frac{x_0}{R_0}\right)^2\frac{eV}{U}\left(\frac{r_y}{R_0}\right)^2\sim10^{-3}\div 10^{-2}.\;\;\;\;\;\;
$$

\section{St\"uckelberg interference in classical two-level system}
Finally, we propose an experimental realization of the spin-mechanical coupling  based on the investigation of the envelope function of vibrating QD' displacements \cite{vdzant}. 
\begin{figure}[t]
\centering
\includegraphics[width=80mm]{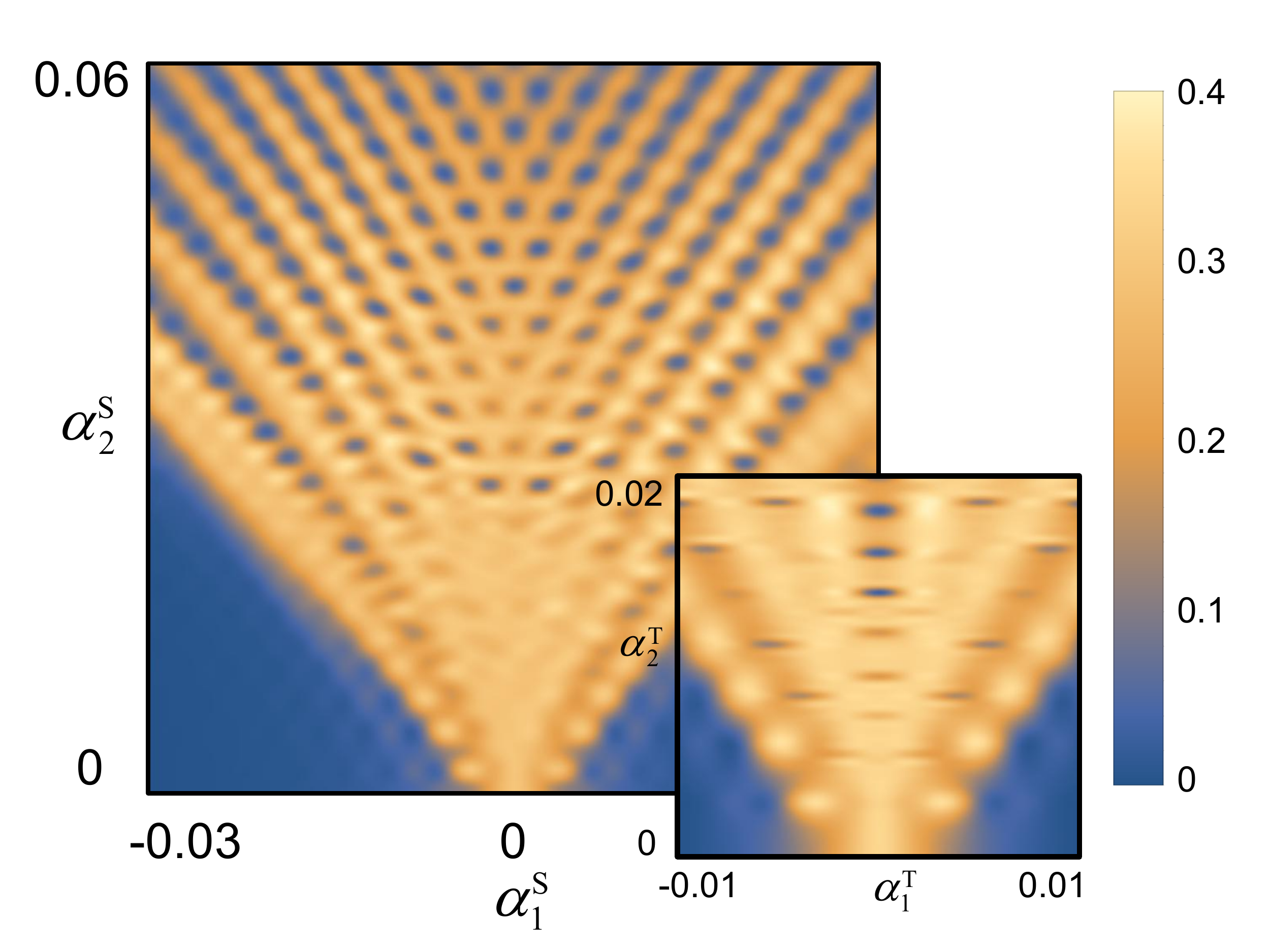}
\caption{(Color online) The time averaged {\color{black} over 30 periods} probability 
{\color{black} $P_{\rm av}$} to populate the in-phase mode  
$|\varphi |^2$ as a function of the dimensionless energy offset (time-independent frequency shift) $\alpha_1$ and the driving amplitude $\alpha_2$. The fanlike diagram is obtained from Eq.(\ref{nm}) for the dimensionless parameters: driving frequency
$\omega_d$$=$$10^{-3}$, detuning $\Delta $$=$$2\cdot 10^{-3}$, and quality factor {\color{black}$Q$$=$$10^5$}. The initial condition for Eq.~(\ref{nm}) assumes
population of the out-of-phase mode at $\tau$$=$$0$: $|\phi(0)|^2$$=$$1$.
Initial velocities of mechanical oscillators are equal to zero. 
We consider the modulations of the density in the back gate corresponding to small deviations of the potential energy around the second node of $V_{\rm eff}$ (Fig. \ref{Fig1}). {\color{black} The two-spin configuration is locked in the singlet state (main panel) and in the triplet state (inset), see detailed discussion about two-spin configuration in Sec. IV}. We neglect the spin-relaxation processes and the effects of hyperfine interaction.} \label{Fig2}
\end{figure}
An idea is based on the observation that the slow dynamics of the NEM system associated with the drive $\omega_d$ mimics the dynamics of a driven quantum two-level system.
The slowly varying amplitudes of in- and out-of-phase modes play the same role as the spinor in the time-dependent Schr\"odinger equation of two-level system \cite{dykhne},\cite{novotny}. To demonstrate the similarity between  classical and quantum driven systems we start with the ansatz \cite{dykhne},\cite{novotny}:
\begin{eqnarray}\label{solution}
\varphi(\tau) &=& C\cdot {\rm Re} \{ \Phi_+(\tau) e^{i\tau-\tau/(2Q)} \},\nonumber\\
\phi(\tau) &=& C\cdot {\rm Re} \{ \Phi_-(\tau) e^{i\tau-\tau/(2Q)} \}.
\end{eqnarray}
The complex amplitudes $\Phi_\pm$ are equivalent to the spinor "wave functions".
Here the constant $C$ accounts for the normalization condition $|\Phi_+|^2$$+$$|\Phi_-|^2 $$\approx $$1$ 
[to achieve normalization condition
we account for the time dependent external force {\color{black}$F$}
in the particular solution of Eq.~(\ref{nm})].
Substituting Eq.~(\ref{solution}) into Eqs.~(\ref{nm}) and performing the unitary transformation with operator {\color{black}$\widehat{W}=e^{i\alpha_1\tau/4}\exp[-i(\alpha_2/4\omega_d)\cos(\omega_d \tau)]$} we map Eqs.~(\ref{nm}) onto a Schr\"odinger-like equation
\begin{eqnarray}\label{sysr}
i\frac{d}{d \tau}\left(
\begin{array}{c}
\Psi_+(\tau)\\
\Psi_-(\tau)
\end{array} \right)=
H_{\rm TLS}(\tau)\left(
\begin{array}{c}
\Psi_+(\tau)\\
\Psi_-(\tau)
\end{array} \right),
\end{eqnarray}
where
\begin{eqnarray}\label{hamil}
H_{\rm TLS}=
-\sigma_x\frac{\Delta}{2}  -\sigma_z\frac{\alpha_1 + \alpha_2\sin(\omega_d \tau)}{4},
\end{eqnarray}
and $\Psi_{\pm}$ are linked to $\Phi_{\pm}$ as {\color{black}$\Psi_{\pm}$$=$$\widehat{W} \Phi_{\pm}$}. 
The instantaneous adiabatic eigenvalues of the Hamiltonian (\ref{hamil}) depend on time $\tau$ as follows $E$$=$$\pm(1/2)\sqrt{\Delta^2+(\alpha_1+\alpha_2\sin(\omega_d \tau))^2/4}$. 
In the vicinity of avoided crossing points, where the distance between two levels is minimal, the linearized model describes the Landau-Zener transitions \cite{LZ} with effective Hamiltonian 
$H_{\rm LZ}$$=$$-(\Delta/2)\sigma_x $$\pm $$(v \tau/2)\sigma_z$, where $v$$=$$\alpha_2\omega_d/2$ {\color{black} is a driving velocity}. 
The Landau-Zener transition occurs between diabatic states associated with in-phase and out-of-phase
modes. As a result, adiabatic states representing the true normal modes of coupled classical oscillators are formed. The transition probability to stay at the same diabatic state after a single passage through the crossing point is given in semiclassical approximation by a text book equation 
\cite{LZ}: $P_{\rm LZ}$$=$$\exp(-\pi \Delta^2/2v)$. 

 In the case of a multipassage process, the transition probability accounts for both diabatic and adiabatic transitions and contains the phase responsible for the interference between two passes \cite{schevch}.
{\color{black} The interference pattern is visualized by a fan-type diagram, see Fig.~\ref{Fig2}. 
The density plot (see Fig.~\ref{Fig2})
shows the time-averaged probability to populate the in-phase mode as a function of the dimensionless "energy offset" (time-independent frequency shift) $\alpha_1$ and the driving amplitude $\alpha_2$.}
The minima and maxima of the time-averaged probability correspond to destructive and constructive interference between consecutive energy levels crossings \cite{footnote3}.
Usually, in the absence of any dissipation, the maximum value of the averaged probability to populate the state satisfying nonoccupied initial condition is equal to 0.5.}
However, the maximum value plotted on Fig.~\ref{Fig2} is below this limit.  To explain 
the probability deficit we point out that the effects of dissipation in a driven nanomechanical system are twofold:
On one hand, these effects invalidate at very large times the correspondence between the full-fledged mechanical equations of motion for the in- and out- of-phase modes and its quantum mechanical equivalent for slowly oscillating amplitudes (envelope curves as "wave functions"); 
on the other hand, the evolution of an
"analogous" two-level system becomes nonunitary. As a result, the maximum value of the averaged probability depends on the number of adiabatic periods used for computing the average value (see Fig.~\ref{Fig2}). 

{\color{black} Specific shape of the fan diagram in Fig.~\ref{Fig2} indicates the crossover from the regime of slow-passage limit $\alpha_2\omega_d\lesssim \Delta^2 $ (bottom part of the figure on the main panel) to the regime of the fast-passage $\alpha_2\omega_d\gtrsim \Delta^2$ (top part of Fig.~\ref{Fig2}). The interference pattern (main panel) demonstrates pronounced arcs similar to Ref. \cite{schevch}. Decrease of the total probability $P_{\rm av}$ with increasing $\alpha_2$ qualitatively reminds the similar effect in the quantum two-level system associated with the presence of two typical times scales of the same order of magnitude responsible for the relaxation and dephasing, see, e.g., \cite{schevch}.}

The standard St\"uckelberg fan diagram \cite{schevch} is constructed assuming mutual independence of parameters $\alpha_1$ and $\alpha_2$. In contrast to it, the RKKY-mediated 
level crossing imposes certain constraint on $\alpha_1$$\sim $$\tilde \alpha_0$ and 
$\alpha_2$$\sim $$\tilde\alpha_0$, making the line $\alpha_1$$=$$0$, $\alpha_2$$\neq $$0$ inaccessible.
The St\"uckelberg interference is pronounced inside the cone $\alpha_1$$<$$\alpha_2$. This condition is achieved by fine tuning the gate voltage and interdot capacitance controlling $U_{12}$.

{\color{black} Finally we comment on relations between two-localized spin configurations 
(singlet/triplet) determined by initial conditions given by RKKY interaction and the resulting St\"uckelberg interference pattern. First,  the spin configuration affects only the magnitude of the coupling constants $\alpha_{1,2}$. 
The expectation value of $\langle{\bf S}_1{\bf S}_2\rangle$ at zero temperature is equal to $-3\hbar^2/4$  for a singlet and $\hbar^2/4$ for a triplet state respectively.
Therefore, for the same values of the external parameters  $\{eV, U, U_{12}\}$, the relation 
$\alpha^{\rm Sing}_{1,2}=-3\alpha^{\rm Trip}_{1,2}$ holds. By constructing the 
interference diagram we assumed very long singlet-triplet relaxation times. Thus,
the system being prepared in certain (singlet or triplet) two-spin initial configuration
is locked in the same state during the evolution. Furthermore, as one can see from
Fig.~\ref{Fig2}, the initial two-spin configuration uniquely defines the  St\"uckelberg
pattern. Therefore, the classical St\"uckelberg interferometry  can be used for the identification of the quantum spin states \cite{footnote4}.}



\section{Summary and discussion}
In summary, we propose {\color{black} a hybrid system coupling two spin impurities embedded in adjacent NEM beams using the RKKY interaction.}
We showed that a nanodevice based on two suspended 
quantum dots nanomachined in the vicinity of a metallic back gate characterized by
a slowly modulated density of charges allows us to control independently both local
and nonlocal spin correlations. The role of the mechanical system is twofold.
On one hand, it provides access to RKKY-mediated dynamics.
On the other hand, it provides a very sensitive tool for quantum measurements of 
nanomechanical back action. The interference 
between two diabatic states of mechanical system
can be measured with high accuracy through  St\"uckelberg oscillations.
The mechanical system of two coupled QD oscillators, while being itself deeply in the classical regime, mimics the dynamics of a quantum two-level system. The slow varying displacement envelope functions play the same role as the two-level system's wave functions \cite{dykhne},\cite{novotny}. 

{ We have demonstrated that the interference between two classical modes (in-phase and out-of-phase) of a mechanical resonator is sensitive to the quantum spin configuration of the double quantum dot. As a result, the St\"uckelberg fan diagram provides very accurate information 
about the spin-spin correlation function. In particular, in the presence of competing interactions, such as, for example, a resonance Kondo scattering, the mechanical back action becomes an important tool for sensing the Kondo screening. The interference pattern, being pronounced for both the singlet and the triplet two-spin configurations, disappears completely when two Kondo clouds are formed in the DQD system to screen the electron's spins. Moreover, mechanical back action can be used to probe the quantum criticality associated with an antagonism between magnetic (RKKY) interaction and Kondo scattering.

The applications of the mobile {\color{black} DQD NEM system} in addition to sensing the spin-spin correlations function include but are not limited by the following problems, to list a few: Competition between resonance on-site Kondo scattering and spin-spin correlation {\color{black} out-of-equilibrium},
nanomechanically induced singlet-triplet transitions in a double-dot device \cite{petta}, 
mechanically induced drag, {\color{black} classical vs quantum} synchronization, etc.} 
{Possible experimental realizations of mechanically tuned RKKY can be engineered with coupled
suspended carbon nanotube/metallic quantum wire resonators or in silicon metal-oxide-semiconductor based junctions, in which mechanics is modeled by driving the barrier gates with an ac voltage \cite{drive}.}

\section*{Acknowledgements}
We are grateful to B. Lorenz and S. Ludwig for fruitful discussions on
St\"uckelberg interference in systems of coupled mechanical resonators,
S. Ilani for many valuable suggestions on possible experimental realizations of quantum nanodevices, F. Pistolesi for critical comments on suspended CNTs, R. Fazio and F. Ludovico for careful reading of the manuscript and valuable comments, Leonid Levitov for drawing our attention to Ref. \cite{dykhne} and
R. Shekhter and L. Gorelik for inspiring discussions on RKKY {\color{black}interaction}. 
This work was finalized at Aspen Center for Physics, which was supported by National Science Foundation Grant No. PHY-1607611 and was partially supported (M.K.) by a grant from the Simons Foundation.


\end{document}